\documentclass[aps,prr,twocolumn,superscriptaddress]{revtex4-2}
\usepackage{amsfonts}
\usepackage{amsmath}
\usepackage{mathrsfs}
\usepackage{amssymb}
\usepackage{graphicx}
\usepackage{lipsum} 
\usepackage[T1]{fontenc}
\usepackage{epstopdf}
\usepackage{bm}
\usepackage{color}
\usepackage{appendix}

\usepackage[colorlinks=true,linkcolor=blue,anchorcolor=blue,citecolor=blue,urlcolor=blue]{hyperref}

\newcommand{\bPsi}{{\bm \Psi}}

\newcommand{\bvarphi}{{\bm \varphi}}

\newcommand{\bw}{{\bm w}}
\newcommand{\bk}{{\bm k}}
\newcommand{\bn}{{\bm n}}
\newcommand{\br}{{\bm r}}

\newcommand{\tmu}{\tilde{\mu}}

\newcommand{\tOmega}{\tilde{\Omega}}

\newcommand{\CH}[1]{\textcolor{red}{{#1}}}

\begin{document}

\title{Transfer of solitons and half-vortex solitons via adiabatic passage}

\author{Chenhui Wang}
% \email{knightwch@outlook.com}
\affiliation{Institute for Quantum Science and Technology, Department of Physics, Shanghai University, Shanghai 200444, China}
\affiliation{
 Centro de F\'{i}sica Te\'orica e Computacional, Faculdade de Ci\^encias, Universidade de Lisboa, Campo Grande, Ed. C8, Lisboa 1749-016, Portugal 
 }

\author{Yongping Zhang}
\email{yongping11@t.shu.edu.cn}
\affiliation{Institute for Quantum Science and Technology, Department of Physics, Shanghai University, Shanghai 200444, China}

\author{V. V. Konotop}
\email{vvkonotop@ciencias.ulisboa.pt}
\affiliation{
 Centro de F\'{i}sica Te\'orica e Computacional, Faculdade de Ci\^encias, Universidade de Lisboa, Campo Grande, Ed. C8, Lisboa 1749-016, Portugal 
 }
 \affiliation{
 Departamento de F\'{i}sica, Faculdade de Ci\^encias, Universidade de Lisboa, Campo Grande, Ed. C8, Lisboa 1749-016, Portugal 
 }

\begin{abstract}

We show that transfer of matter-wave solitons and half-vortex solitons in a spin-orbit coupled Bose-Einstein condensate between two (or more) arbitrarily chosen sites of an optical lattice can be implemented using the adiabatic passage. The underlying linear Hamiltonian has a flat band in its spectrum, so that even sufficiently weak inter-atomic interactions can sustain well-localized Wannier solitons which are involved in the transfer process. The adiabatic passage is assisted by properly chosen spatial and temporal modulations of the Rabi frequency.  Within the framework of a few-mode approximation, the mechanism is enabled by a dark state created by coupling the initial and target low-energy solitons with a high-energy extended Bloch state, like in the conventional stimulated Raman adiabatic passage used for the coherent control of quantum states. In real space, however, the atomic transfer between initial and target states is sustained by the current carried by the extended Bloch state which remains populated during the whole process. The full description of the transfer is provided by the Gross-Pitaevskii equation.  Protocols for the adiabatic passage are described for one-- and two--dimensional optical lattices, as well as for splitting and subsequent transfer of an initial wavepacket simultaneously to two different target locations.  
\end{abstract}

\maketitle

\section{Introduction} 

Localization and transfer of matter or energy between different states of a system with periodically varying parameters is one of the central problems across various branches of physics, with particular relevance to Bose-Einstein condensates (BECs)~\cite{Brazhnyi2004,Bloch2005,Morsch2006} and optics~\cite{Kartashov2011}. In a defect-free medium, the localization is achieved using nonlinearity, which in one-dimensional (1D) systems enables gap solitons. Although at small-amplitudes such solitons can move through a lattice, they are wide envelopes of Bloch states extending over dozens of the lattice periods. Much stronger localization at weak nonlinearities can be achieved in systems featuring flat bands. In particular, in specially designed discrete systems, perfect localization may be sustained by strictly flat bands even in the linear regime~\cite{Leykam2018,Poblete2021}. In continuous periodic systems, band flatness can only be achieved  approximately. Nevertheless, if a band of the underlying linear spectrum is nearly flat, then even extremely weak nonlinearity is sufficient for sustaining Wannier solitons (WSs)  which remain localized on a single period of the potential, as discussed in~\cite{AKKS} and~\cite{Wang2023} for one- and two-component BECs, respectively. A drawback of such localization is complete suppression of the soliton mobility. Thus, localization and controlled mobility of gap solitons cannot be achieved simultaneously without employing auxiliary physical mechanisms. 
  
In this paper, we describe a mechanism enabling the transfer of strongly localized states of a spin-orbit coupled (SOC) BEC~\cite{Lin2011, Galitski2013, Wu2016, Sun2018, Wang2021, Hamner2015, Hamner2014} between two (or more) {\it a priori} chosen distant sites in one-dimensional (1D) and two-dimensional (2D) optical lattices (OLs). Specifically, we extend the ideas of the stimulated Raman adiabatic passage (STIRAP)~\cite{Bergmann1998, Vitanov2017, Bergmann2019} known to be a highly efficient tool for selective transfer between quantum states of different energy levels, and in several situations, between spatially confined eigenstates, such as states in a three-well parabolic trap in both linear~\cite{Eckert2006} and nonlinear~\cite{Graefe2006} regimes, to {\em spatial} transfer of wavepackets in extended nonlinear systems.  

We note that the STIRAP principle has already been explored for the spatial transfer of energy between separated linear edge states in a 1D array of optical waveguides in a topologically nontrivial phase~\cite{Longhi2019}. Our goal here is to use the adiabatic passage for implementing and controlling spatial transfer of matter WSs and half-vortex ({\em alias} semi-vortex) solitons~\cite{Lobanov2014, Malomed2019} in 1D and 2D OLs respectively, when both initial and target states are located in prescribed arbitrarily chosen lattice sites. The distinguishing features of the phenomenon reported here include the confinement of the states due to weak nonlinear inter-atomic interactions, mediated by the flat band of the underlying linear Hamiltonian spectrum, without requiring any specific topology of the system. Furthermore, the reported adiabatic passage allows for the splitting of the original soliton and the delivery of atoms to two (or more) different target solitons.

The paper is organized as follows. In Sec.~\ref{sec:model}, we formulate the model that governs the adiabatic passage and discuss physical parameters. The few-mode approximation for a SOC-BEC in a periodic potential with a flat band is elaborated in Sec.~\ref{few_mode}.  Numerical simulations of one-to-one transfer are presented in Sec.~\ref{1D} and Sec.~\ref{2D} for 1D and 2D OLs, respectively.  In Sec.~\ref{Multi-target}, we extend the ideas of the adiabatic passage to transfer the initial single WS (or two WSs) into two target solitons at different locations. \CH{The effect of the type of nonlinearity is addressed in Sec.~\ref{nonlinearity},} and a brief summary of the main outcomes is given in the Conclusion.

\section{The model}
\label{sec:model}

Consider a $D$-dimensional two-component SOC-BEC with attractive two-body interactions, which is loaded in an OL. In the mean-field approximation, the system is described by the spinor ${\bPsi}=\left({\Psi}_{1},{\Psi}_{2}\right)^{T}$ governed by the dimensionless Gross-Pitaevskii equation (GPE) 
\begin{align}
    \label{GPE}
    i\partial_t\bPsi=H_D\bPsi+\Omega(\br,t)\sigma_z\bPsi+G(\bPsi^\dag,\bPsi)\bPsi,
\end{align}
where
\begin{align}
\label{H_1D}
  H_1=-\frac{1}{2} \partial_x^{2}-i\gamma\sigma_{y}\partial_x 
+V_0\sin^{2}x  
\end{align}
for $D=1$ and 
\begin{align}
\label{H_2D}
 H_2=-\frac{1}{2}(\partial_x^2+\partial_y^2)
-i\gamma\left(\sigma_{y}\partial_x -\sigma_{x}\partial_y\right)
\nonumber \\
+V_{0}(\sin^2x+\sin^2y)   
\end{align}
for $D=2$ are linear Hamiltonians, $\gamma$ is the SOC strength, $V_{0}$ is the amplitude of the respective OL, 
$\Omega(\br,t)=\Omega_0+ \tilde{\Omega}(\br,t)$ is the Rabi frequency  whose deviation $\tilde{\Omega}(\br,t)$ from the consant value $\Omega_0$ is to be established latter on, and $\sigma_{x,y,z}$ are the Pauli matrices.
The nonlinearity
\begin{align}
    G =\left(\begin{array}{cc}
     g_{11}\left|{\Psi}_{1}\right|^2+g_{12}\left|{\Psi}_{2}\right|^2    & 0  \\
       0  & g_{12}\left|{\Psi}_{1}\right|^2+g_{22}\left|{\Psi}_{2}\right|^2
    \end{array}\right)
\end{align}
originates from the atomic interactions characterized by the coefficients $g_{ij}$.

In the dimensionless GPE (\ref{GPE}) the units of the energy, Rabi frequency and potential depth are measured in the units of $2E_{L}$, where $E_L=\hbar^{2}k_{L}^{2}/2m$ is the recoil energy of the OL, $m$ is the atomic mass and $k_L$ is the wave vector of the lattice beams projected onto the longitudinal direction. The spatial coordinate is measured in the units of $1/k_L$. The effective SOC $\gamma= k_R/k_L$, where $k_R$ is the wave number of the Raman beams in the longitudinal direction can be varied either by changing the angle between the incident Raman beams or through fast modulation of the intensities of the Raman lasers~\cite{experiment3, Zhang2013_1}.
The quantities $\Omega(\br,t)$ and $V_0$, which can be tuned by changing the intensities of the Raman and lattice beams, respectively. 
The orders of magnitude of the dimensionless quantities involved, can be estimated using typical experimental settings with  $^{87}$Rb atoms~\cite{Lin2011, Galitski2013, Hamner2015, Wu2016, Sun2018, Wang2021, Hamner2014} with transverse linear oscillator frequencies $\omega_{y,z} =2\pi\times 150\,$Hz,  $k_L\approx 3\mu\, {\rm m}^{-1}$ and $k_R\approx 6\,\mu{\rm m}^{-1}$, yielding $E_L\approx3.5\times10^{-31}$J. Then $\gamma\approx 2$, while the Rabi frequency $\Omega_0$ and the depth of the lattice potential $V_0$ in Eq.~(\ref{GPE}) explored below vary over a range extending up to a dozen recoil energies. 

The strength and sign of inter-atomic interactions $g_{ij}$ can be changed by means of the Feshbach resonance~\cite{Chin2010}. The adiabatic passage reported here is not too sensitive to specific choice of the nonlinear interactions, since it is based on the optimization procedure involving several parameters. Nevertheless, it is more robust and efficient for attractive interactions. Considering also that typically absolute values of scattering lengths $a_{jj}$ of inter-species interactions is smaller than that of intra-species interactions, most of our numerical results reported below are shown for the choice $g_{11}=g_{22}=-1$, $g_{12}=-0.1$. This choice of $g_{jj}$ implies the normalization of the dimensionless order parameter expressed by: $N=\int d\bm{r}\bPsi^\dagger\bPsi$ such that $\mathcal{N}=N\mathcal{N}_D$, with $\mathcal{N}_1=E_L\sqrt{\omega_y\omega_z}/(\hbar k_{L}|a_{11}|)$,
$\mathcal{N}_2=E_{L}\sqrt{2\pi\hbar\omega_{z}/m}/(\hbar k_{L}^{2}|a_{11}|)$, gives the physical numbers of atoms for $D=1$ and $D=2$, respectively. 
For typical values of the parameters used below $\mathcal{N}_1\sim 1200$, and $\mathcal{N}_2\sim 200$. 

\begin{figure}[t]
	\includegraphics[width=\columnwidth]{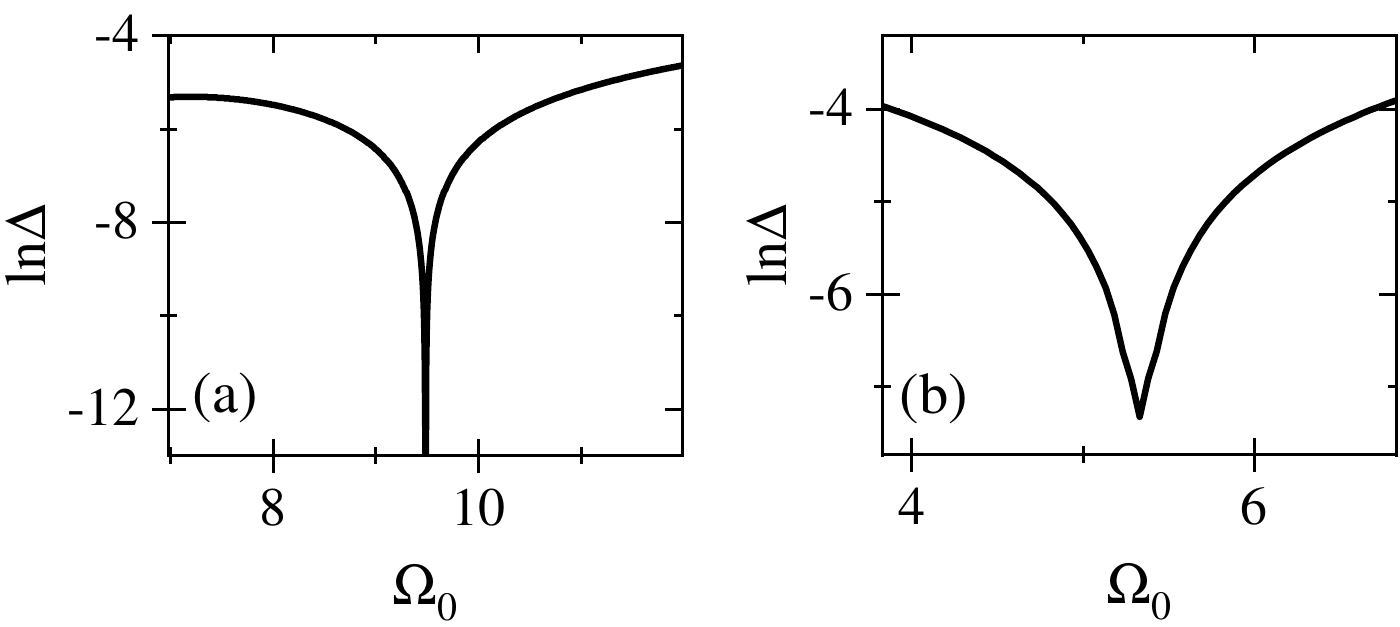}
	\caption{The flatness $\Delta$ versus the constant Rabi frequency $\Omega_0$ for (a) the 1D system with $\gamma=2.5$, $V_0=3$ and (b) the 2D system with $\gamma=2$, $V_0=2$. The minimal   flatness is $\Delta=1.02\times10^{-6}$ in (a) and $\Delta=6.58\times 10^{-4}$ in (b).}
        \label{fig:one}
\end{figure}
The chemical potential $\mu_\nu(\bk)$ of the linear SOC-BEC with a constant Rabi frequency $\Omega_0$ is obtained from
\begin{align}
    (H_D+\Omega_0\sigma_z)\bvarphi_{\nu\bk}=\mu_\nu(\bk)\bvarphi_{\nu \bk}.
\end{align}
Here, $\bvarphi_{\nu \bk}$ is the Bloch spinor (it can be obtained numerically by the plane-wave expansion~\cite{Hui2017}), $\nu$ is the band index, and  $\bk=k$ and $\bk=(k_x,k_y)$ in the 1D and 2D cases, respectively. We are interested in the case when the lowest band, $\nu=0$, is nearly flat~\cite{Zhang2013, Zhang2015, Kartashov2016a, Kartashov2016b, Hui2017, Wang2023},  the flatness being characterized by the band width~\cite{Wang2023} 
\begin{align}
    \Delta =\max_k \mu_0(\bk)-\min_k\mu_0(\bk).
\end{align}
Figure~\ref{fig:one} illustrates that the parameters $\gamma$ and $\Omega_0$ can indeed be chosen to achieve extreme flatness with $\Delta \sim 10^{-6}$ and $\Delta\sim 10^{-4}$ in 1D and  2D cases as shown, respectively [see also Fig.~\ref{fig:two} (a), below]. 

While in a linear OL, no localized states can exist, due to flatness of a band, even very weak nonlinearity can support WSs that can be created at arbitrarily chosen lattice sites. In either dimension, profiles of such solitons are well approximated by Wannier functions (WFs)~\cite{Kohn1959} [for flat-band WSs in a 1D SOC-BEC see~\cite{Wang2023}, while in a 2D case they are discussed below]. 
However, an "instant" flatness of the lowest band determined at $t=0$ varies over the time because of adiabatic variation of the system parameters. This raises an additional (technically) challenging problem of finding such SOC variations, i.e., the function $\tilde{\Omega}(\br,t)$, which does not affect the localization of the involved WSs because of increase of the bandwidth  and at the same time ensures the adiabatic passage protocol described below and schematically illustrated in Fig.~\ref{fig:two} (a).

\begin{figure}[t]
	\includegraphics[width=\columnwidth]{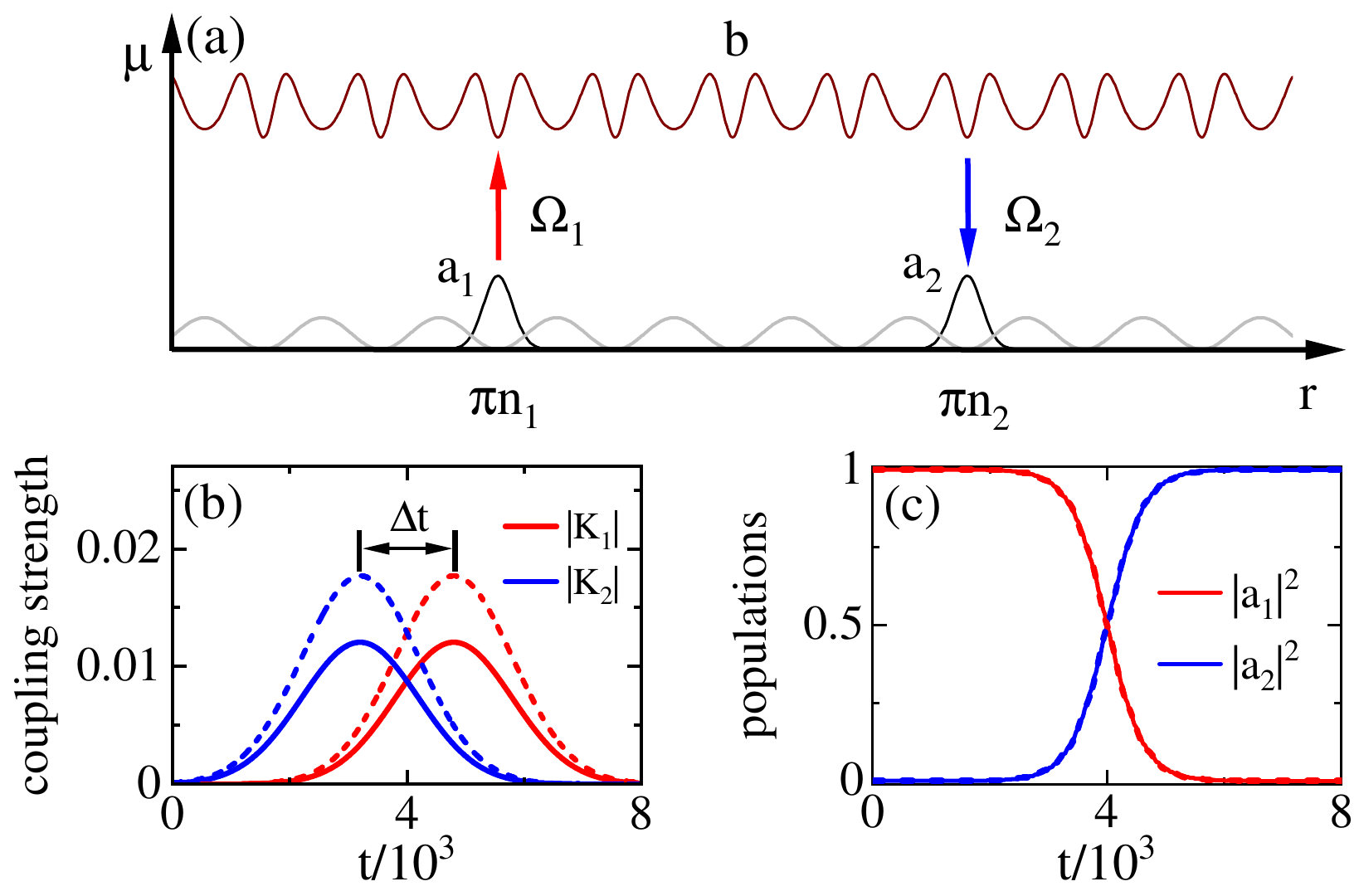}
    \caption{(a) Schematics of the one-to-one soliton transfer via adiabatic passage in 1D case (2D case is similar). The initial WS, the target WS, and the extended state are marked by $a_1$, $a_2$, and $b$, respectively. $\Omega_2$ and $\Omega_1$ are the Rabi couplings embody the first and second stages of the protocol [$\Omega_1$ is delayed with respect to $\Omega_2$, as shown in panel (b)]. The light-gray line shows the OL.  (b) Time evolution of the coupling strengths $K_i(t)$ defined in (\ref{K}).  (c) Evolution of the populations of the initial state (red lines) and the final state (blue lines) obtained from the three-mode approximation (see the text). Solid and dashed lines in (b) and (c) are for the 1D and 2D cases, respectively. Here and below, unless specified otherwise, we set $\gamma=2.5$, $\Omega_0=9.482$, $V_0=3$, $k=-0.42$, $\varpi=2.25$, $\alpha=0.9$ for 1D case, and  $\gamma=2$, $\Omega_0=5.331$, $V_0=2$, $\bk=(-1/2,-1/2)$, $\varpi=1.5$, $\alpha=1$ for 2D case. In both cases $\Delta t=1600$, $\beta=5\times10^{-7}$, $N=0.01$.}
        \label{fig:two}
\end{figure}

In a relatively shallow lattice  only the lowest band can be flat [for 2D case it is illustrated in Fig.~\ref{fig:four} (a)]. The goal of the adiabatic passage can be formulated as follows. Having created a flat-band WS with a relatively small amplitude $a_1$ at a lattice site $\bn_1$ [hereafter ${\bm n}=n$ in 1D case and ${\bm n}=(n_x,n_y)$ in 2D case; $n$ and $n_{x,y}$ are integers], one must transfer the atoms to other prescribed lattices sites $\bn_{2,3,...,M}\neq \bn_1$. To implements such transfer one applies  modulations of the Rabi frequencies $\tOmega_{2,3,...,M}$ in the vicinity of the target spatial locations which at the same time couples the flat band with an extended Bloch state of a certain amplitude $b$ in the energy space. With some time delay $\Delta t$ (or delays $\Delta t_{2,3,...,M}$) modulation of the Rabi frequency $\tOmega_1$ localized in space around ${\bm n}_1$ is applied. Due to temporal overlap of the modulations $\tOmega_1$ and $\tOmega_{2,3,...,M}$ the atomic transfer from the extended state to target WSs with the amplitudes $a_{2,3,...,M}$ at the indicated locations is fulfilled enables by the extended Bloch state. Below we distinguish {\em one-to-one} [see schematics in Fig.~\ref{fig:two} (a)] and {\em multi-target} transfers depending whether the prescribed new location consists of the only or several lattice sites, respectively.

The outlined approach suggests exploring modulations of the the Rabi frequency  of the form $\tilde{\Omega}=\varpi \cos(\omega t) \sum_{k=1}^M \tilde{\Omega}_k$, where $\omega=\mu_b-\mu_0$ with $\mu_b$ being the chemical potential of the excited Bloch sate, while specific shapes of the local couplings can be chosen as Gaussians: $\tilde{\Omega}_j= \exp[{-\alpha(\br-\pi \bn_j)^2-\beta(t-t_j)^2}]$. The free control parameters $\varpi$, $\alpha$, $\beta$, and $\Delta t_{j}=t_{1}-t_{j}>0$  are to be chosen to optimize the transfer, i.e., to achieve the smallest loss of atoms in the fastest transfer process.

The adiabatic passage scheme, cannot produce the target WS identical to the original one because of unavoidable particle losses occurring during the process. In our dimensionless nonlinear problem, for the attractive nonlinearity, the norm $N$ that parameterizes soliton families (and determines the number of atoms) is a decreasing function of the chemical potential $\mu$. 
%[see e.g.  $N(\mu)$ in Fig.~\ref{fig:three} (d) below] 
Hence, the chemical potentials of the original, $\mu_{\rm init}$, and target, $\mu_{\rm targ}$, WSs are not equal, but $\mu_{\rm init}<\mu_{\rm targ}$. Hence, $N(\mu_{\rm targ})$ is not known {\em a priori}. This requires some assumptions for engineering $\tilde{\Omega}(\br,t)$. The assumption adopted here is that the chemical potentials of the initial and target solitons are nearly equal: $\mu_{\rm init}\approx \mu_{\rm targ}$, what is justified when the transfer has the efficiency high enough.    
 
\section{A few-mode approximation}
\label{few_mode}

The simplest estimates for the parameters of the multi-target (and one-to-one) transfer can be obtained using the $M$-mode approximation when the leading order solution of the GPE is searched in the form~\cite{Wang2023}
\begin{align}
\label{ansatz}
\bPsi= e^{-i\mu_{0}t}\sum_{j=1}^{M}a_{j}(t)\bw_{\bn_{j}}(\br)+e^{-i\mu_{b}t}b(t)\bvarphi_{b}(\br),
\end{align}
where $a_{1}(t)$, $a_{2,...,M}(t)$, and $b(t)$ are time-dependent amplitudes of the initial WS, target WSs and of the (quasi-linear) Bloch state, respectively. 
First, we consider the case of one-to-one transfer corresponding to $M=2$. Projecting over the Wannier functions $\bw_{\bn_{1,2}}$ and over the Bloch function $\bvarphi_{b}(\br)$ we obtain
\begin{align}
\label{dyn-eq}
    i\frac{d{a}_{1,2}}{dt}=K_{1,2}b, \qquad i\frac{d{b}}{dt}=K_1^*a_1 +K_2^*a_2
\end{align}  
where 
\begin{align}
\label{K}
    K_{j}(t)=\frac{\varpi}{2}\int_{\mathbb{R}^{D}}\tilde{\Omega}_{j}\bw_{\bn_{j}}^{\dagger}  \sigma_{z}\bvarphi_{b}d\br
\end{align}
are the coupling strengths, an asterisk means complex conjugation (see Appendix.~\ref{appendix_threemode} for more details). These equations conserve the total number of atoms: $\left|a_{1}\right|^2+\left|a_{2}\right|^2+\left|b\right|^2=N$. 

Since the goal is to transfer atoms from the initial state $\bPsi_{\rm in}\propto \bw_{\bn_1}$ to the target state $\bPsi_{\rm tg}\propto \bw_{\bn_2}$ avoiding the transient population of the state $\bvarphi_{b}$, we address the dark-state ($b\equiv 0$): 
\begin{align}
\label{dark}
    \bPsi_{\rm dark}=\frac{K_2^*\bw_{\bm{n}_{1}}-K_1^*\bw_{\bm{n}_{2}}}{\sqrt{\left|K_{1}\right|^2+\left|K_{2}\right|^2}}.
\end{align}
The model (\ref{dyn-eq}) formally coincides with that of the conventional STIRAP process~\cite{Bergmann2019, Merkel2007, Nesterenko2009, McEndoo2010, Greentree2004, Vitanov2017}. Using this analogy, for an optimized scheme we require $|K_1|\ll|K_2|$ ($|K_2|\ll|K_1|$) during the initial up (final down) transitions, as well as the condition $\int_{-\infty}^{\infty}\left(\left|K_{1}\right|^2+\left|K_{2}\right|^2\right)^{1/2} dt\gg 1$ ensuring sufficient energy gap between the dark state and the two excited states~\cite{Bergmann1998}. Using the control parameters $\varpi$, $\alpha_{1,2}$ and $\beta$ one can satisfy these constraints and design the evolution of the coupling coefficients as shown in Fig.~\ref{fig:two} (b) for 1D and 2D cases. 

To obtain WFs numerically, following the procedure described in ~\cite{Marzari2012}, we computed a smooth gauge $\phi(\bk)$ and the sufficiently well-localized WFs $\bw_{\bn_{j}}$ of the flat band by projection, which are defined as 
\begin{align}
\label{eq_wannier}
    \bw_{\bn_j}(\bm{r})&=\frac{1}{2^D}\int_{BZ} d\bm{k}\bvarphi_{0 \bm{k}}e^{i\phi(\bk)-i\pi\bm{k}\cdot\bm{n_j}},
\end{align}
where the integration is over the first Brillouin zone.  
In the 1D case, we set the system size $30\pi$ with $2^{13}$ grid points, while in the 2D system, the space size is $10\pi\times10\pi$ with $2^{9}\times2^{9}$ grid points. Meanwhile, periodic boundary conditions are imposed, and varying the space size can eliminate the influence of boundary conditions.
An example of 2D WF is shown in Fig.~\ref{fig:two}(b-c). 

 \begin{figure}[t]
	\includegraphics[width=\columnwidth]{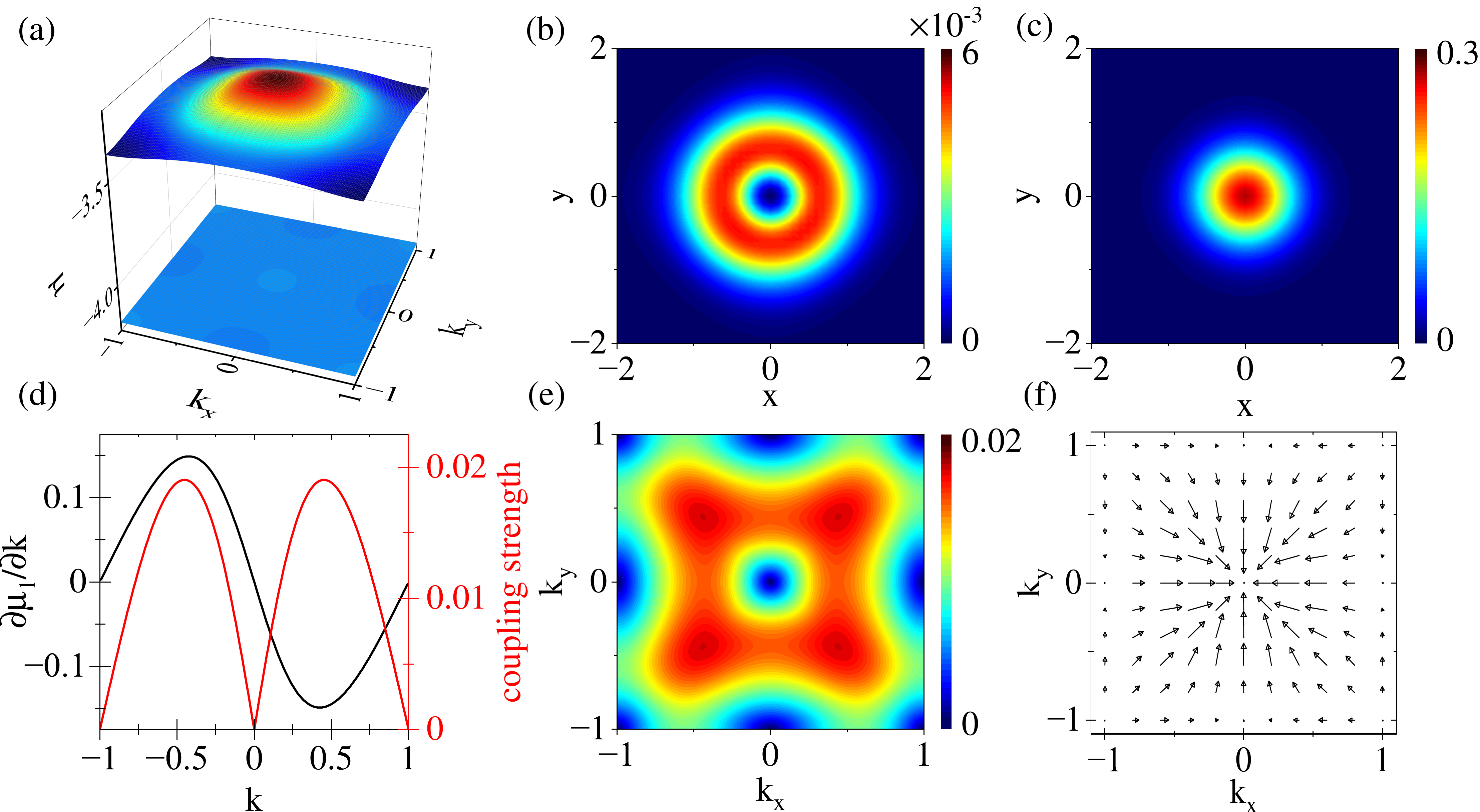}
	\caption{(a) Two lowest bands of the 2D linear Hamiltonian with parameters from Fig.~\ref{fig:one}(b). The lowest band at $\mu_0\approx-4.186$ is flat: $\Delta=6.59\times10^{-4}$ (its landscape is non-distinguishable on the plot scale). The first and second components of a 2D WF are shown in (b) and (c), respectively. (d) Coupling strength $K_i$ vs $\bk$ of the Bloch state from the upper branch for (d) 1D system (red line) and (e) 2D system. The black line in panel (a) shows the group velocity ${\bm v}_{\bk}$ in the 1D system. For the 2D case, the absolute value of group velocity $|{\bm v}_{\bk}|$ and its directions {\it versus} $\bk$ are shown by arrows in panel (f). The parameters are (a-c), (e-f) $\gamma=2.5, \Omega_0=9.482, V_0=3$, and (d) $\gamma=2, \Omega_0=5.331, V_0=2$. 
   } 
        \label{fig:two}
\end{figure}

So far, we treated the extended state formally. The specific choice of $\bvarphi_b(\br)$ is made to ensure that the Bloch state carries atoms between initial and target WSs with the largest group velocity for the first (non-flat) band $\nu=1$ ,  This requirement reduces the time during which the excited state has nonzero population. Respectively, $\bk$ is found from maximization of the group velocity  
${\bm v}_{\bk}=\partial \mu_1(\bk)/\partial \bk$
modulus  $|{\bm v}_{\bk}|$. 
In Figs.~\ref{fig:two} (d) and (e)  we show the  coupling strengths $K_j$ for the 1D and 2D cases.
The respective group velocities ${\bm v}_{\bk}$  are shown in Figs.~\ref{fig:two} (d) and (f). One observes that the maximum coupling strength $K_i$, indeed appears around the maximum of the modulus of group velocity $|{\bm v}_{\bk}|$ at $k\simeq-0.42$ in the 1D case [Fig.~\ref{fig:two} (d)] and near $\bk=(-1/2,-1/2)$ in the 2D case [Figs.~\ref{fig:two} (e) and (d)]. 

\section{Adiabatic passage in real space: 1D case}
\label{1D}

\begin{figure}[t]
	\includegraphics[width=\columnwidth]{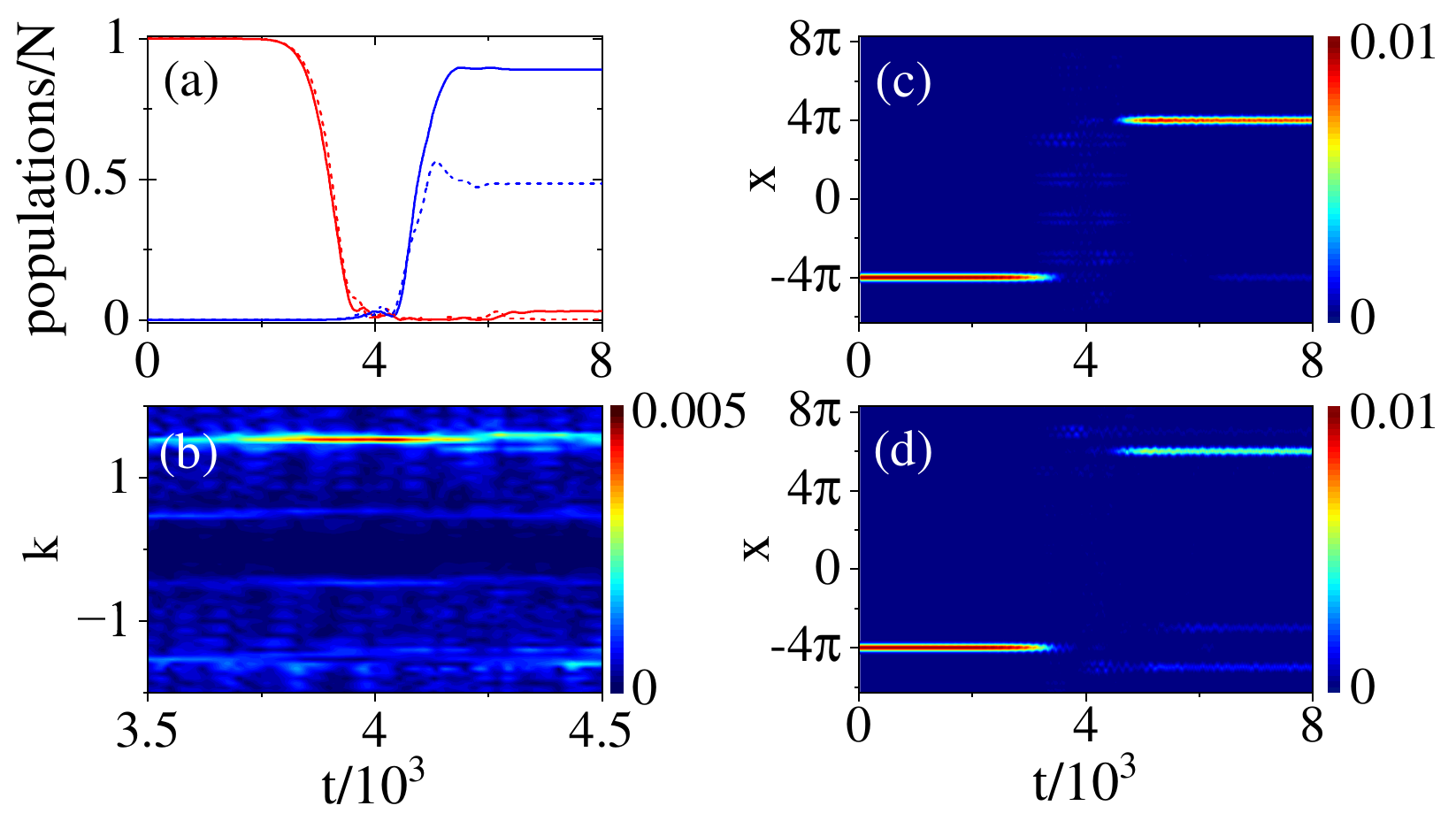}
	\caption{(a) Evolution of relative populations $|a_1|^2/N$ (red lines) and $|a_2|^2/N$ (blue lines) obtained from the 1D GPE (\ref{GPE}) for initial WS located at $\bn_1=-4$ and target WSs located at $\bn_2=4$ (solid lines) and at $\bn_2=6$ (dashed lines) The evolution of the intensity distributions in these two cases are shown in (c) and in (d), respectively. (b) Evolution of the spatial Fourier spectrum corresponding to panel (c). The most populated Fourier mode has $k\approx 1.57$, corresponding to the Bloch mode with $k\approx -0.43$ in the reduced BZ scheme.  Here, $\omega=1.299$, $t_1=4600$, $t_2=3400$, $\alpha=0.9$, and $\bPsi(t=0)=\sqrt{N}\bw_{\bn_1}$.
  }
        \label{fig:four}
\end{figure}

Unlike in the conventional STIRAP process, the use of a few-mode model for spatially extended systems, like the one considered here, has several additional limitations. First, either the amplitude of the Rabi frequency $\varpi$, or the free control parameter $1/\beta$ can be scaled out in the few-mode model, what is impossible to do in the original GPE~(\ref{GPE}). 
Second, the coupling strength $|K_j|$ defined in (\ref{K}) is independent of the position of the maximum of $|\bw_{\bn_{j}}|$, i.e., the vector $\pi\bn_j$ is not accounted for by Eq.~(\ref{dyn-eq}), while the GPE accounts for these factors which are important for dispersion and instabilities that can be observed in a continuous system. Third, the few-mode approach does not account for variations of the band flatness $\Delta_0$ occurring due to deviation of the Rabi frequency from the constant value $\Omega_0$. Indeed, Fig.~\ref{fig:one} shows the flatness $\Delta$ vs the Rabi frequency $\Omega_0$ in 1D and 2D cases, respectively [see also Fig.~\ref{fig:four} (a)]. Thus, on the one hand, to avoid significant increase of the band width some limitations on the maximal value of the $\varpi$ must be imposed. On the other hand, as $\varpi$ increases, the coupling strength $K_i$, which is proportioning to $\varpi$, enhances the coupling between states and promotes the soliton transfer. Finally, the dimensionality of the space affects only the magnitudes of the coupling strengths $K_{1,2}$ in the few-mode model, while for the original GPE it is a crucial factor. Therefore, next we proceed to study of the adiabatic passage described by Eq.~(\ref{GPE}). Meantime, the results obtained with using the few-mode model will be employed as initial guess for the parameters optimising the transfer of atoms in real space. 

Results of numerical simulation for the one-to-one transfer are reported in Fig.~\ref{fig:four}, where we explore the parameters of the adiabatic passage shown in Fig.~\ref{fig:two} (c) by solid lines. One observes that the transfer efficiency can achieve almost $87\%$ when the distance between initial and target solitons is 8 periods [Fig.~\ref{fig:four} (a) and (c)], while for the same parameters of the system it reduces to only $48\%$ for the distance of 10 periods [Fig.~\ref{fig:four} (a) and (c)]. The difference of efficiencies is natural considering the longer time of the transfer, and thus larger loss of atoms (due to exciting undesirable extend states) as well as due to dipersion effect. On the other hand, the result in  Fig.~\ref{fig:four} (d) are not optimal [the parameters wer optimized for the transfer shown in Fig.~\ref{fig:four} (c)] and can improved by further adjusting the parameters. The panels (c) and (d) show spatial distributions of the atomic densities in these processes, which justifies the above assumption about the proximity of the respective chemical potentials (especially in case of $8$-period distance). 

In the real process, one can control only the Rabi frequency, but cannot "chose" the extended Bloch state $\bvarphi_b$ directly. In Fig.~\ref{fig:four} (b) we show the evolution of the Fourier spectrum of the excited mode during the adiabatic passage (only the interval during which the population of the extended Bloch state is maximal is shown).  We observe that the excited states remain weakly populated at all instants of the transfer process (note the data shown in the color bar): the major number of the atoms is concentrated in the Bloch state with $k\approx -0.43$, what agrees well with the prediction of the model (\ref{ansatz}), (\ref{dyn-eq}) yielding $k\approx -0.42$ [see Fig.~\ref{fig:two} (d)]. 
 
\section{Adiabatic passage in 2D case}
\label{2D}

\begin{figure}[t]
	\includegraphics[width=\columnwidth]{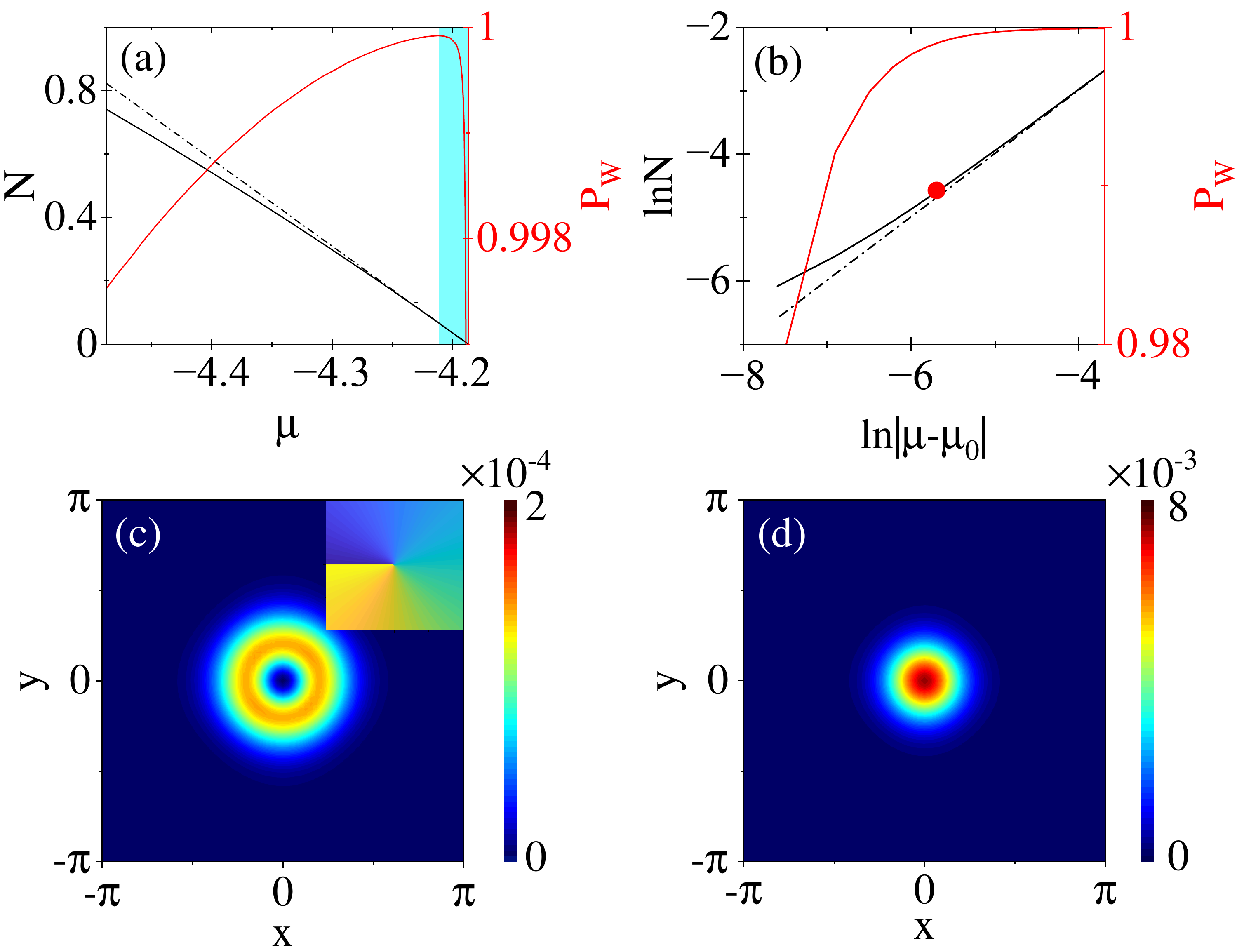}
	\caption{(a,b) Families of 2D half-vortex solitons (black lines) and the projection $P_W$ (red lines) on the 2D WF are shown in the interval $\mu<\mu_0$. In panel (a), the flat band is located on the right side (indicated by the light-cyan shadow area). The ln-ln plot in panel (b) zooms in the region in the vicinity of the flat band $\mu_0$ (in this plot the linear band is located at $-\infty$ of the abscissa). In panels (a,b), solid lines are for the numerical results of Eq.~\ref{GPE}, and dashed lines are for the results obtained from the approximation $\mu-\tmu_0=\chi_0N$. The first and second components of a half-vortex soliton with $N=0.01$ and $P_W=0.999$ are shown in (c) and (d), respectively, for the points marked by the red dot in panel (b). The inset in (c) shows the phase of the first component. The parameters are $\gamma=2, \Omega_0=5.331, V_0=2, \chi_0=0.365$.}
        \label{fig:five}
\end{figure}

Now we turn to adiabatic passage of a SOC-BEC loaded in a 2D square OL described by the Hamiltonian (\ref{H_2D}). We consider transfer of the flat-band 2D half-vortex soliton [the linear spectrum and the respective Wannier function are illustrated in Fig.~\ref{fig:two} (a)-(c)].
By analogy with the 1D case, near the flat-band edge, half-vortex solitons remain localized on the scale of one lattice period, having shapes that can be approximated by 2D WFs. 
An example of a small amplitude half-vortex soliton carrying phase singularity in only one component is shown in Figs.~\ref{fig:five}~(c) and ~\ref{fig:five}~(d) [cf. Figs.~\ref{fig:two}~(b) and ~\ref{fig:two}~(c); note however that the Wannier function is normalized to one, while the norm of the shown soliton is $N=0.01$. 

The whole family of 2D half-vortex soliton is characterized by the linear dependence 
\begin{align}
\label{mu-N}
    \mu-\tmu_0=\chi_0N
\end{align}
obtained by small-amplitude expansion, where 
\begin{align}
\label{chi_0}
    \chi_0=\int_{\mathbb{R}^{D}} d\br\bw_{0}^\dagger G(\bw_0^\dagger, \bw_0)\bw_0.
\end{align}
The details of the derivation are given in Appendix.~\ref{appendix_expansion}, while in Fig.~\ref{fig:five} (a), (b), we show the numerical confirmation of the dependence (\ref{mu-N}), (\ref{chi_0}). In Fig.~\ref{fig:five} (a), (b), we show the projection of a 2D half-vortex soliton on the 2D WF: 
\begin{align}
\label{PW}
    P_W=\int_{\mathbb{R}^{D}} d\br\bw_0^\dagger\bPsi,
\end{align}
remain practically unchanged in a wide interval of the chemical potentials, even in the proximity of the flat band.

Several distinguished features of the transfer in the 2D case, compared to the 1D case considered in the previous section, are to be mentioned. Stronger dispersion of the excited Bloch mode mediating the adiabatic passage, generally speaking, may greatly reduce its efficiency, where the direction of the atom transfer is prescribed by the direction of the group velocity. Two-dimensional half-vortex solitons are particularly vulnerable to instabilities now including possibilities of collapse. Additionally, it is known that instability of vortices can be developed upon tunneling through a double-well potential~\cite{Salgueiro2009} or when the vortex is subject to coherent tunneling adiabatic passage~\cite{McEndoo2010} even with a short transfer distance. 

Thus, we first checked the stability of half-vortex solitons by both the linear stability analysis of the respective Bogoliubov–de Gennes equations and by direct long-time propagation governed by the GPE (\ref{GPE}), (\ref{H_2D}) with the addition of initial Gaussian noise of order of $5\%$ of the soliton amplitude. Both approaches manifested consistent results, indicating that the family of half-vortex solitons is found stable in the whole region of the parameters where the approximation (\ref{mu-N}) holds. 

% \begin{figure}[t]
% 	\includegraphics[width=\columnwidth]{Fig_wannier.pdf}
% 	\caption{(a,b) The Wannier functions profiles in 2D Rashba spin-orbit coupling BEC. (c)  The projection $P_{W}$ of the fundamental WS on the WF $\bw_0$ vs. $\mu-\mu_0$ and (d) its zoom (shown using the logarithmic scale) in the proximity of the flat band [light-cyan shadow area in the panel (c)]. }
%         \label{fig:wannier}
% \end{figure}
 
\begin{figure}[t]
	\includegraphics[width=\columnwidth]{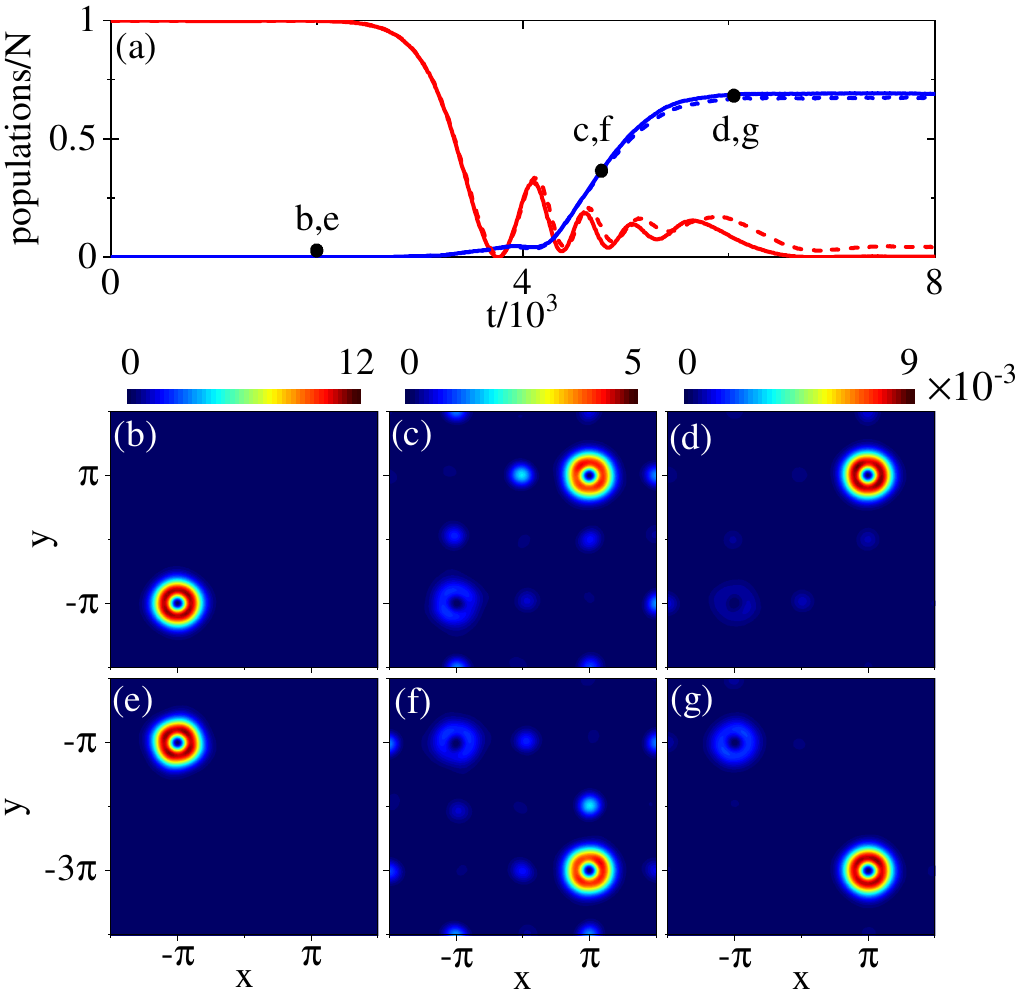}
	\caption{(a) Relative populations the initial $|a_1|^2/N$ (red lines) and final $|a_2|^2/N$ (blue lines) of half-vortex solitons, obtained from (\ref{GPE}). The initial state is prepared at $\bm{n}_{1}=(-1,-1)$ while the target solitons are at $\bn_{2}=(1,1)$ [panels (b)-(d); solid lines in (a)]  and at $\bn_{2}=(1,-3)$ [panels (e)-(g); dashed lines in (a)]. The snapshots in panels (b) to (g) show only the first components having phase singularity at the instants indicated in (a). Here, $\omega=0.889$, $t_1=4800$, $t_2=3200$, $\alpha=1$ and $\bPsi(t=0)=\sqrt{N}\bw_{\bn_1}$.}
        \label{fig:six}
\end{figure}

For the numerical study of the adiabatic passage of half-vortex solitons, we obtained the optimized parameters for the Rabi frequency using (\ref{ansatz}), (\ref{dyn-eq}). The results for the configuration corresponding to one shown in Fig.~\ref{fig:six} are depicted in Fig.~\ref{fig:two} (b) and (c) by dashed lines. The found parameters are used for exploring (numerically) the adiabatic passage within the framework of GPE (\ref{GPE}). Two characteristic scenarios are shown in Fig.~\ref{fig:six} for the transfer of a half-vortex soliton in different spatial directions but over equal distances [cf. panels (b)-(d) with (e)-(g)].  
In Fig.~\ref{fig:six} (a), where populations of the initial and target half-vortex solitons are shown, one observes that the total transfer rates for both processes are approximately equal to $69\%$ (note that these cases are not linked by any symmetry).

\section{Multi-target adiabatic transfer}
\label{Multi-target}

Turning to {\em multi-target} transfer through adiabatic passage, the full optimization problem becomes significantly more complex. In the meantime, it is relatively simple to present examples as proof of concept that the described mechanism works in such a statement, too. First, we consider atoms initially constituting a soliton located at $\bn_1$, which are split into two groups and delivered to different target locations $\bn_2$ and $\bn_3$. Thus, now $M=3$ in (\ref{ansatz}), (\ref{dyn-eq}). We consider equal delays $\Delta t_2=\Delta t_3$, keeping other parameters the same as Fig.~\ref{fig:two}. 

\begin{figure}[t]
	\includegraphics[width=\columnwidth]{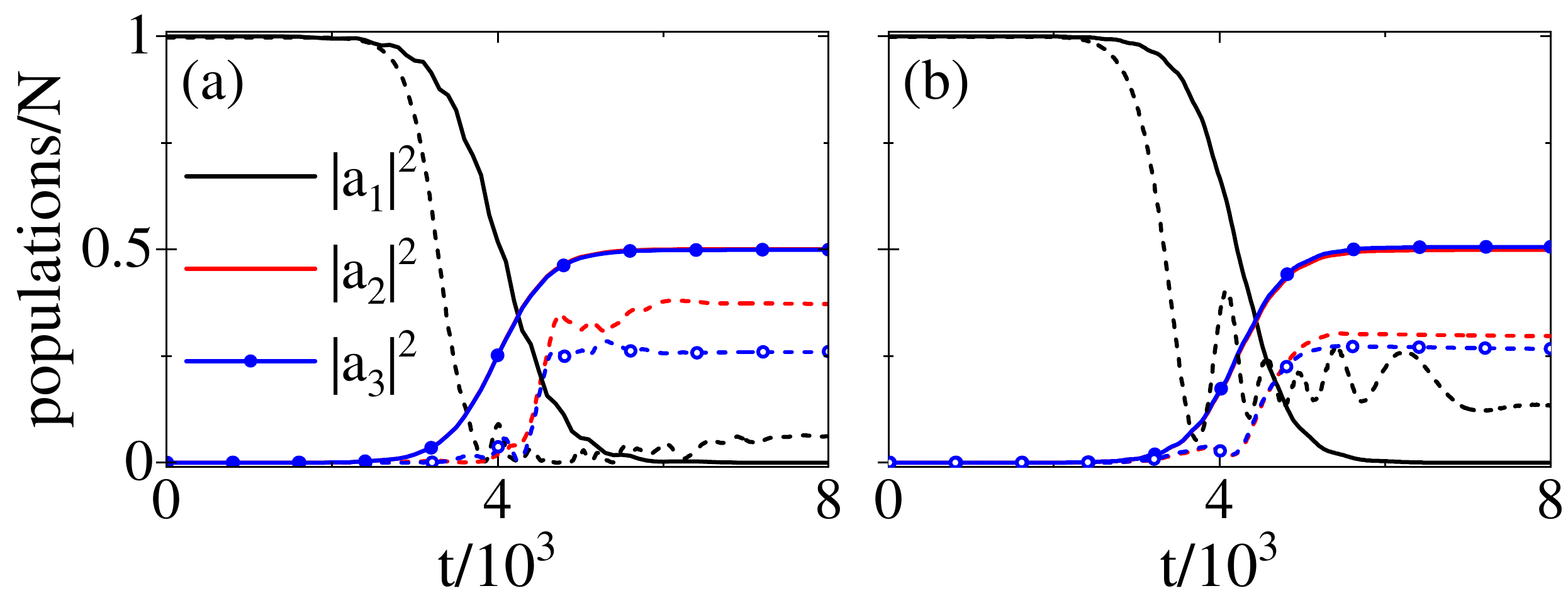}
	\caption{Multi-targets transfer of WSs obtained numerically from the four-mode model (solid lines) and Eq.~(\ref{GPE}) (dashed lines). In (a) 1D case is shown for $n_1=-4$, $n_2=4$ and $n_3=6$. In (b) 2D case is shown for $\bn_1=(-1,-1)$, $\bn_{2}=(1,1)$, and $\bn_3=(1,-3)$. Other parameters are the same as Fig.~\ref{fig:four} and Fig.~\ref{fig:six} for (a) and (b), respectively.}
        \label{fig:seven}
\end{figure}

Examples of this adiabatic passage in 1D and 2D cases are illustrated in Figs.~\ref{fig:seven} (a) and (b), respectively. In the 1D case, when the population of the initial WS is transferred to the positions $4\pi$ and $6\pi$, which separately were considered in Fig.~\ref{fig:four} (b) and (d), the respective efficiencies are $32\%$ and  $26\%$ of atoms. Thus, at the end of the adiabatic passage, the lattice site closer to the initially prepared state has a higher population than the more distant site. For the splitting and adiabatic passage of half-vortex solitons shown in Fig.~\ref{fig:seven}, the efficiencies are $29\%$ and $28\%$ for the target solitons at $\bn_2$ and $\bn_3$, respectively. 

\begin{figure}[t]
	\includegraphics[width=\columnwidth]{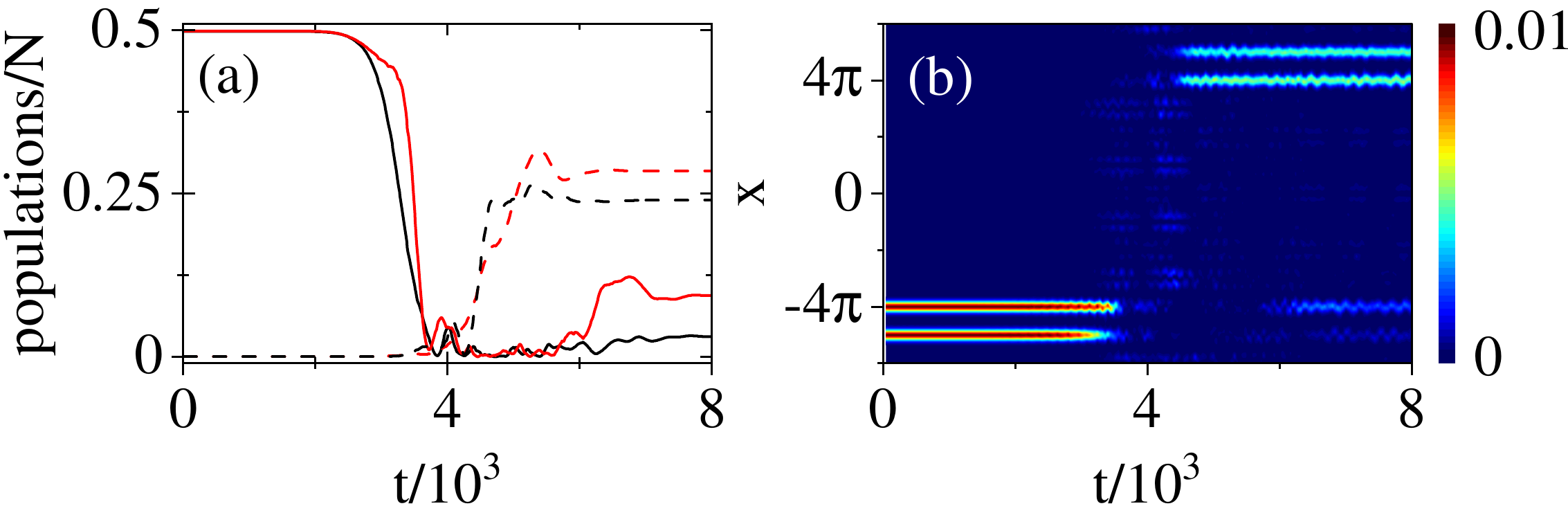}
	\caption{(a) Two-to-two transfer of WSs obtained numerically from Eq.~(\ref{GPE}), which is shown for $n_1=-5, n_2=-4$ (black and red solid lines), and $n_3=5, n_4=4$ (black and red dashed lines).   The corresponding evolution of the intensity distributions is shown in panel (b). Here, $N=0.02, \bPsi(t=0)=\sqrt{N/2}(\bw_{\bn_1}+\bw_{\bn_2})$, Other parameters are the same as Fig.~\ref{fig:four}.}
        \label{fig:nine}
\end{figure}

Remarkably, we find that if the initial pulse is a multi-hump soliton~\cite{Wang2023}, i.e., if initially atoms populate several lattice sites, the multi-target transfer adiabatic passage still works. In Fig.~\ref{fig:nine} we show an example a 1D {\em two-to-two} transfer of atoms. Specifically, atoms of two WSs created at adjacent lattice sites $n_1=-4\pi$ and $n_2=-5\pi$, are transferred to two adjacent target locations $n_3=4\pi$ and $n_4=5\pi$.  We observe the achieved efficiencies are $28\%$ (at $n_3$) and $24\%$ (at $n_4$). As mentioned above, now the optimization procedure becomes too complicated because of multiple parameters involved. Furthermore, it is not uniquely defined because the final goal and optimization costs allow for different formulations (e.g., higher efficiency at a desired location, such is $n_3$ in the shown case, although it could also be $n_4$). Therefore the results represent a proof of principle rather than the best protocol for two-to-two transfer. 

\begin{figure}[t]
	\includegraphics[width=\columnwidth]{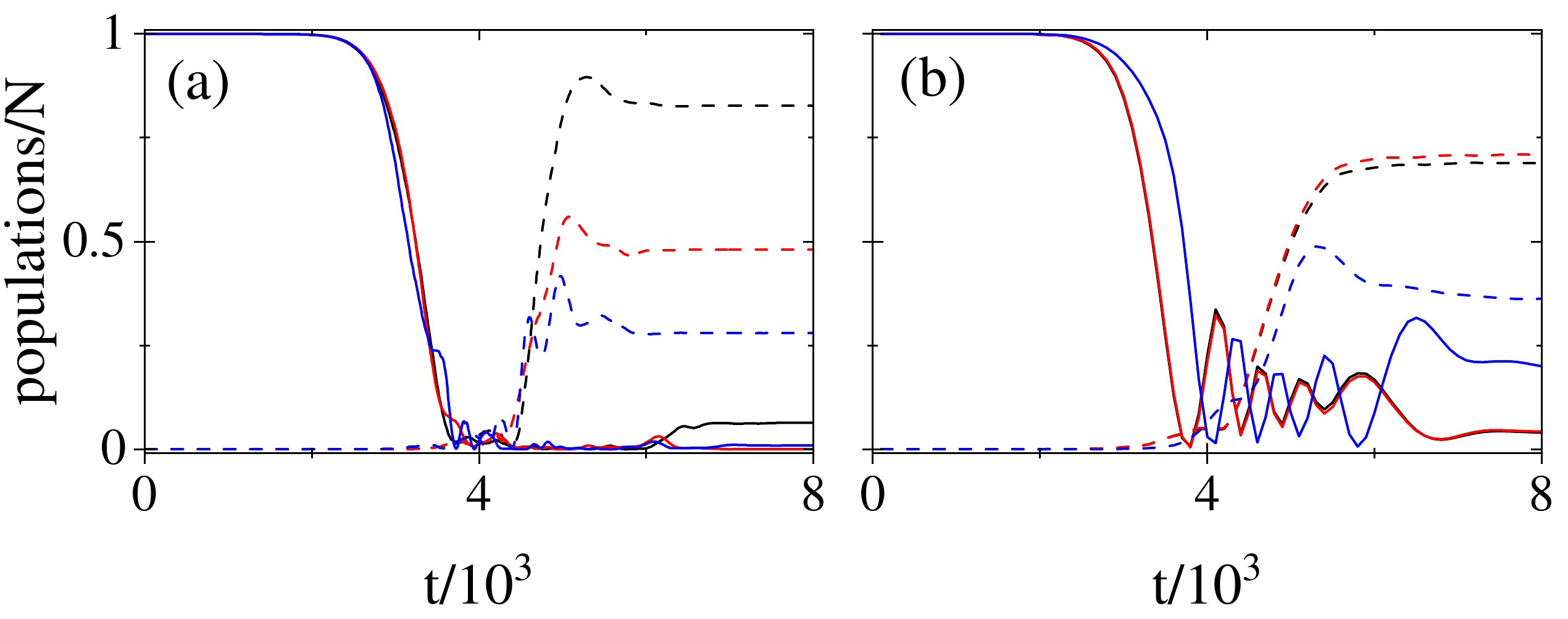}
	\caption{Relative populations the initial $|a_1|^2/N$ (solid lines) and final $|a_2|^2/N$ (dashed lines) of the adiabatic passage for (a) 1D WSs and (b) 2D half-vortex solitons obtained numerically from Eq.~(\ref{GPE}). In panel (a), the initial location is $n_1=-4$, while $n_2=4$ and $g_{ij}=-1$ (black lines),   $n_2=6$ and $g_{ij}=-1$ (red lines), and $n_2=4$ and $g_{11}=g_{12}=1$, $g_{22}=0.9954$ (blue lines). In panel (b), the initial location is $\bn_1=(-1,-1)$, while $\bn_2=(1,1)$ and $g_{ij}=-1$ (black lines), $\bn_2=(1,-3)$ and $g_{ij}=-1$ (red lines), and  $\bn_2=(1,1)$ and $g_{11}=g_{12}=1$, $g_{22}=0.9954$ (blue lines). Other parameters in (a) and (b) are the same as Fig.~\ref{fig:four} and Fig.~\ref{fig:six}, respectively.}
        \label{fig:eight}
\end{figure}

\section{Different nonlinearity}
\label{nonlinearity}

Finally, we briefly address the effect of the type of nonlinearity for adiabatic passage in a spinor-type system. While in the preceding sections, we set the attractive interaction as $g_{11}=g_{22}=-1, g_{12}=-0.1$,  the magnitudes of $g_{ij}$ can be chosen within a relatively wide range. Furthermore, in the more common situation of a repulsive condensate $g_{ij}$ are positive. Remarkably, in this latter case, the adiabatic passage still works reasonably well for finite distances, although the conversion rate is lower than in the case of negative scattering lengths. Examples of one-to-one transfer for all positive scattering lengths are shown by blue lines in Figs.~\ref{fig:eight} (a) and ~\ref{fig:eight} (b) for the 1D and 2D cases, respectively. One observes that the  repulsive nonlinearity makes the pulses localized and tends to suppress the adiabatic passage leading to lower efficiencies, as compared with all attractive interactions; cf. blue lines obtained for the experimentally available condensate of $^{87}$Rb atoms~\cite{Hamner2015,Sun2018} and black lines obtained for all equal attractive interactions. In the repulsive case the achieved efficiencies are $27\%$ and $37\%$ shown by dashed lines in Figs.~\ref{fig:eight} (a) and ~\ref{fig:eight} (b), respectively. The higher efficiency in the 2D case [panel (b)] is attributed to much smaller distance $2\sqrt{2}\pi$ between the initial and target states, while the same distance is $8\pi$ in the 1D case [panel (b)]. Meantime the black and red lines, showing transfer over the distances of 8 and 10 periods, show the  efficiencies $82\%$ and $48\%$, respectively,   These numerical results indicate that our protocol is relatively robust with respect to variations of the nonlinear attractive interactions and can be experimentally implemented in repulsive condensates, as well.

\section{Conclusion}
\label{conclusion}

To conclude we have described how an adiabatic passage can be used for the spatial transfer of low-amplitude matter solitons in spin-orbit-coupled BEC loaded in an optical lattice, whose linear spectrum features a flat band. The protocols were elaborated for the transfer between two solitons at {\it a priori} prescribed locations in one- and two-dimensional settings, as well as for simultaneous split of the initial wavepacket into two ones with subsequent transfer of the parts to different target locations. The physical mechanism based on the use of the dark state coupling two (or more) nonlinear localized wavepackets with a higher energy extended Bloch state is generic and can be employed in a variety of physical media sustaining simultaneously spatially localized states (solitons) and delocalized states. For instance, similar protocols could be used to implement transfer through the adiabatic passage in optical waveguides and photorefractive crystals, as well as for structured acoustic media.

\acknowledgments
The work was supported by the Portuguese Foundation for Science and Technology (FCT) under Contracts UIDB/00618/2020 (DOI: 10.54499/UIDB/00618/2020) and PTDC/FIS-OUT/3882/2020 (DOI: 10.54499/PTDC/FIS-OUT/3882/2020), and by the China Scholarship Council (CSC) under the Grant No. CSC N.202206890002. This work was also supported by the National Natural Science Foundation of China (NSFC) with Grants No. 12374247 and No. 11974235, and by the Shanghai Municipal Science and Technology Major Project (Grant No. 2019SHZDZX01-ZX04).

\appendix
\section{Derivation of the three-mode model}
\label{appendix_threemode}

To the derive equations of the three-mode approximation we use the three-mode ansatz (\ref{ansatz})  with $M=2$.
\begin{align}
\label{ansatz_app}
 \bPsi&= e^{-i\mu_{0}t}\sum_{j=1}^{2}a_{j}(t)\bw_{\bn_{j}}(\br)
 \notag\\
 &+ b(t)\int_{BZ}d\bk e^{-i\mu_{1}(\bk)t} s (\bk)\bvarphi_{1 \bk}(\br),
\end{align}
where $a_{1,2}(t)$ and $b(t)$ are slowly varying functions of time, $s (\bk)$ is the spectrum of the excited Bloch states $\bvarphi_{1 \bk}$, and only the leading terms of the expansions are used~\cite{Wang2023}. 

Notice that the Wannier and Bloch functions are normalized as follows
\begin{align}
    \label{Normal}
    \int_{\mathbb{R}^{D}}d\br\bw_{n_i}^\dagger\bw_{n_j}&=\delta_{ij},
    \\ \int_{\mathbb{R}^{D}}d\br\bvarphi_{\nu' \bk'}^\dagger\bvarphi_{\nu \bk}&=2^D\delta_{\nu\nu'}\delta(\bk-\bk'),
\end{align} 
while normalization of the Bloch wavepacket requires
\begin{align}
    \label{norm-s}
    \int_{BZ}d\bk |s(\bk)|^2 =1/2^D.
\end{align}

Substituting Eq. (\ref{ansatz_app}) into the equation of linear SOC-BEC: $i\partial_{t}\bPsi=[H_{D}+\Omega(\br,t)\sigma_z]\bPsi$ and keeping only the leading order terms~\cite{Wang2023} we obtain
\begin{align}
\label{aux0}
 i\sum_{j=1}^{2}\frac{da_{j}}{dt} \bw_{\bn_{j}}+i\frac{db}{dt}  
     \int_{BZ}d\bk e^{-i[\mu_{1}(\bk)-\mu_{0}]t} s(\bk) \bvarphi_{1 \bk}
       \notag\\
       =\varpi \sigma_{z} \cos(\omega t) (\tOmega_1+\tOmega_2)
       \nonumber \\ 
       \times \left[\sum_{j=1}^{2}a_{j}\bw_{n_{j}}+  
     b\int_{BZ}d\bk e^{-i[\mu_{1}(\bk)-\mu_{0}]t}  s(\bk) \bvarphi_{1 \bk}\right].
\end{align}

Projecting this equation on $\bw_{\bn_{1,2}}$ and using the orthogonality of the Wannier function of the flat band to Bloch states of other bands, we obtain
\begin{align}
\label{aux1}
     &i\frac{da_{1,2}}{dt} =\varpi  \cos(\omega t)\Big\{ \sum_{j=1}^{2} a_{j}\int_{\mathbb{R}^{D}}  d\br (\tOmega_1+\tOmega_2)\bw_{\bn_{1,2}}^{\dagger}\sigma_{z}  
       \bw_{n_{j}}
       \nonumber \\
       % +& \varpi  \cos(\omega t)
       &+b \int_{\mathbb{R}^{D}}  d\br (\tOmega_1+\tOmega_2) 
     % \notag\\
     %   \times&
       \int_{BZ}d\bk e^{-i(\mu_{1}-\mu_{0})t}  s(\bk)\bw_{\bn_{1,2}}^{\dagger}\sigma_{z} \bvarphi_{1 \bk}\Big\},
\end{align}
Averaging over the fast period $2\pi/\omega$, i.e., using the rotating wave approximation (RWA), 
% one eliminates the first term in right-hand side of the above equation, as well as the term having the frequency $\mu_{1}(\bk)-\mu_{0}\pm \omega$ except $\mu_{\nu_0}(\bk)-\mu_{0}- \omega=\mu_{\nu_0}(\bk)-\mu_{b}$ where $\nu_0$ stands for the number of the band to which $\mu_b$ belongs 
(recall that $\omega=\mu_b-\mu_0>0$). 
Thus, denoting
\begin{align}
\label{phi_b}
    \varphi_b=\int_{BZ}d\bk e^{-i[\mu_{1}(\bk)-\mu_{b}]t} s (\bk)\bvarphi_{1 \bk},
\end{align}
Eq.~(\ref{aux1}) is reduced to 
\begin{align}
\label{aux2}
     i\frac{da_{1,2}}{dt}=\frac{\varpi}{2}b \int_{\mathbb{R}^{D}}  d\br (\tOmega_1+\tOmega_2)\bw_{\bn_{1,2}}^{\dagger}\sigma_{z}\varphi_b.
\end{align}
 
Now we take into account that $\tOmega_j$ is localized in the same cell as $\bw_j$, i.e., the condition
\begin{align}
\label{local}
    \left|\int_{\mathbb{R}^{D}}d\br \bw_{\bn_{j}}^{\dagger}\tilde{\Omega}_{3-j}\sigma_{z}\varphi_b 
    \right|
    \ll \left|\int_{\mathbb{R}^{D}}d\br \bw_{\bn_{j}}^{\dagger}\tilde{\Omega}_{j}\sigma_{z}\varphi_b \right|,
\end{align}
is satisfied, and reduce (\ref{aux2}) to the form $(j=1,2)$
\begin{align} 
\label{aux3}
     i\frac{da_{j}}{dt} = \frac{\varpi}{2} b   \int_{\mathbb{R}^{D}}  d\br\tOmega_j \bw_{\bn_{1,2}}^{\dagger}\sigma_{z}\varphi_b.
\end{align}

Projecting (\ref{aux0}) on $\bvarphi_{1 \bk}$ we obtain
\begin{align}
\label{aux4}
    i 2^Ds(\bk)\frac{d b}{d t}  
       =\varpi \cos(\omega t)e^{i\mu_{1}(\bk)t}
       \notag\\
       \times \int_{\mathbb{R}^{D}} d\br(\tOmega_1+\tOmega_2)\bigg[e^{-i\mu_{0}t}\bvarphi_{1 \bk}^\dagger\sigma_z\sum_j a_{j}\bw_{n_{j}}
       % \left(a_{1}\bvarphi_{1 \bk}^\dagger\bw_{n_{1}}+ a_{2}\bvarphi_{1 \bk}^\dagger\bw_{n_{2}} \right)
       \nonumber \\
       +b\int_{BZ}d\bk'e^{-i\mu_{1}(\bk')t}s(\bk')\bvarphi_{1 \bk}^\dagger\sigma_z\bvarphi_{1 \bk'}\bigg]. 
\end{align}
Now, excluding a very special case of resonances, i..e, assuming that $\mu_{1}(\bk')-\mu_{1}(\bk)\neq \omega$, the second term in the square brackets of this expansion is eliminated in the RWA.  Multiplying Eq.~(\ref{aux4}) by $s(\bk)^*$, using (\ref{norm-s}), (\ref{phi_b}), (\ref{local}), and integrating over the first BZ, we obtain
\begin{align}
\label{aux5}
   i \frac{d b}{d t}=   \frac{\varpi}{2}\int_{BZ} d\bk e^{i[\mu_1(\bk)-\mu_{0}]t}s^*(\bk)\int_{\mathbb{R}^{D}} d\br 
   \notag \\  \times 
   \sum_ja_{j}\tOmega_j \bvarphi_{1 \bk}^\dagger\sigma_z\bw_{n_{j}} 
   \notag\\
   =\frac{\varpi}{2}\int_{\mathbb{R}^{D}} d\br \left(a_{1}\tOmega_1\bvarphi_{b}^\dagger\sigma_z\bw_{n_{1}}+a_{2}\tOmega_2 \bvarphi_{b}^\dagger\sigma_z\bw_{n_{2}} \right).
\end{align}

Finally, from (\ref{aux3}) and (\ref{aux5}), we obtain (\ref{dyn-eq}), (\ref{K}). In the approximation of a sufficiently narrow spectral width of the wavepacket, desirable for the resonant three-wave process $s(\bk)\to \delta (\bk-\bk_b)$ where $\bk_b$ is obtained from the exact resonance condition $\mu_b=\mu_{1}(\bk_b)$, we recover $\bvarphi_b\propto e^{i\bk_b\cdot\br}$ used in the main text.

\section{Small-amplitude expansion}
\label{appendix_expansion}
Recall that in the linear limit, WFs $\bw_{\bn}$ are not eigenstates,
together with Eq.~(\ref{GPE}) and Eq.~(\ref{eq_wannier}), one can obtain
\begin{align}
\label{HW}
		(H_D+\Omega_0\sigma_z)\bw_{\bn}
  =&\frac{1}{2^D}\int_{BZ}d\bk \mu_0(\bk)\bvarphi_{0 \bk}e^{-i\pi\bm{k}\cdot\bm{n}}
		\nonumber \\
  % =&\mu_{0}\bw_{\bn}+\sum_{\bm m}\frac{1}{2^D}\int_{BZ} \left[\mu_0(\bk)-\mu_0\right]\bw_m e^{i\pi\bm{k}\cdot(\bm{m}-\bm{n})}d\bk
		% \nonumber \\
          =&\tmu_{0}\bw_{\bn}+\sum_{\bm m\neq \bn}f_{\bn-\bm m}\bw_{\bm m},
\end{align} 
where $ \tmu_{0}=\mu_0+f_0,$ and
\begin{align*}
     f_{\bm m}=\frac{1}{2^D}\int_{BZ}d\bk \left[\mu_0(\bk)-\mu_0\right] e^{-i\pi\bm{k}\cdot\bm{m}}.
\end{align*}

While in the 1D case, families of WSs bifurcate from the flat band, in the 2D case, this is impossible. However, in this last case, the family can very closely approach the band (which is impossible in the case of non-flat bands), and the expansion for the soliton family in this case works similarly to that in the 1D case. Respectively, using that all $f_{\bm m}$ are nearly zero (zero would correspond to an ideally flat band) and  assuming (without loss of generality) that a soliton family is centered at $\br = 0$, we look for a solution of the GP equation in the form of the “mixed” (Wannier for the lowest band, and Bloch for all other bands) expansion:
\begin{align}
	\label{c-expan}
	\bPsi= \sqrt{\epsilon}e^{-i\tmu t} \left[a(t) \bw_0+\epsilon  \bPsi_1+\mathcal{O}(\epsilon^2)\right],
\end{align}
where
\begin{align}
\label{psi1}
	\bPsi_1= \sum_{\bn\neq 0} b_{\bn}(t)\bw_{\bn}+\sum_{\nu\neq0} \int_{BZ}d\bk b_\nu (\bk,t)\bvarphi_{\nu \bk}.
\end{align}

By substituting the above expansions in the GPE~(\ref{eq_wannier}), we obtain
\begin{align}
\label{complete}
	i\frac{d a}{d t}\bw_0+i\epsilon\sum_{\bn\neq 0} \frac{d b_{\bn}}{d t} \bw_n+i\epsilon \sum_{\nu\neq0} \int_{BZ}d\bk  \frac{\partial b_\nu}{\partial t} \bvarphi_{\nu \bk} 
 \notag	\\
 = a\sum_{\bm{m}\neq0}f_{\bm{m}}\bw_{-\bm{m}}+\epsilon\sum_{\bn\neq 0}b_{\bn}\sum_{\bm{m}\neq 0}f_{\bm{m}}\bw_{\bn-\bm{m}}
 \nonumber \\
 + \epsilon \sum_{\nu\neq0} \int_{BZ}d\bk b_\nu[\mu_{\nu}(\bk)-\tmu_0]\bvarphi_{\nu \bk}\notag\\
 -\epsilon |a|^2aG(\bw_0^\dagger,\bw_0)\bw_0+\mathcal{O}(\epsilon^2).
\end{align}
Projecting on the WFs $\bw_0$, $\bw_{\bn}$ (with $\bn\neq0$) and the Bloch states $\bvarphi_{\nu \bk}$ within the accepted accuracy we obtain, respectively
\begin{align}
    i \frac{d a}{d t} 
    =& \epsilon \chi_0|a|^2a+\epsilon \sum_{n\neq 0}  b_{\bn} f_{\bn},\\
    \epsilon i \frac{d b_{\bn}}{d t} =& af_{-\bn} +\epsilon\sum_{\bm{m}\neq 0}  b_{\bm{m}}f_{\bm{m}-\bm{n}} +\epsilon \chi_{\bm{n}} |a|^2a,\\
    i\frac{\partial b_\nu}{\partial t}=&b_\nu[\mu_{\nu}(\bk)-\tmu_0]+\frac{1}{2^D}\chi_{\nu \bk} |a|^2a,
\end{align}
where
\begin{align}
   \chi_{\bn}&=\int_{\mathbb{R}^{D}}d\br\bw_{\bn}^\dagger G(\bw_0^\dagger,\bw_0)\bw_0 , \\ 
   \chi_{\nu\bk}&=\int_{\mathbb{R}^{D}}d\br\bvarphi_{\nu\bk}^\dagger G(\bw_0^\dagger,\bw_0)\bw_0.
\end{align}
The expansion (\ref{c-expan}) does not allow for the formal limit $\epsilon\rightarrow0$ because $\bw_0$  is not an eigenstate of the linear Hamiltonian.
Since we are interested in a flat band, we estimate $\max\{ f_{\bn}\}\sim\Delta\ll1$. Then, in the leading order, we can neglect all $f_{\bn}$ and obtain (\ref{mu-N}).


\begin{thebibliography}{100}

\bibitem{Brazhnyi2004}
V. A. Brazhnyi and V. V. Konotop, Theory of nonlinear matter waves in optical lattices, \href{https://www.worldscientific.com/doi/abs/10.1142/S0217984904007190}{Mod. Phys. Lett. B {\bf 18}, 627 (2004)}.

\bibitem{Bloch2005} I. Bloch, Ultracold quantum gases in optical lattices, \href{https://doi.org/10.1038/nphys138}  
{Nat. Phys. {\bf 1}, 23 (2005)}.  

\bibitem{Morsch2006} 
O. Morsch and M. Oberthaler, Dynamics of Bose-Einstein condensates in optical lattices, \href{https://journals.aps.org/rmp/abstract/10.1103/RevModPhys.78.179}{Rev. Mod. Phys. {\bf 78}, 179 (2006)}.

\bibitem{Kartashov2011} 
Y. V. Kartashov, B. A. Malomed, and L. Torner, Solitons in nonlinear lattices, \href{https://journals.aps.org/rmp/abstract/10.1103/RevModPhys.83.247}{Rev. Mod. Phys. {\bf 83}, 247 (2011)}.

\bibitem{Leykam2018}
D. Leykam, A. Andreanov, S. Flach, Artificial flat band systems: from lattice models to experiments, \href{https://www.tandfonline.com/doi/full/10.1080/23746149.2018.1473052}{Adv. Phys. X {\bf~3}, 1473052 (2018)}.

\bibitem{Poblete2021}
R. A. V. Poblete, Photonic flat band dynamics, \href{https://www.tandfonline.com/doi/full/10.1080/23746149.2021.1878057}{Adv. Phys. X {\bf~6}, 1878057 (2021)}.

\bibitem{AKKS} 
G. L. Alfimov, P. G. Kevrekidis, V. V. Konotop, and M. Salerno, Wannier functions analysis of the nonlinear Schro\"{o}dinger equation with a periodic potential, \href{https://journals.aps.org/pre/abstract/10.1103/PhysRevE.66.046608}{Phys. Rev. E {\bf 66}, 046608 (2002)}.

\bibitem{Wang2023}
C. Wang, Y. Zhang, and V. V. Konotop, Wannier solitons in spin-orbit-coupled Bose-Einstein condensates in optical lattices with a flat band,
\href{https://journals.aps.org/pra/abstract/10.1103/PhysRevA.108.013307}{Phys. Rev. A {\bf~108}, 013307 (2023)}.

\bibitem{Lin2011}
Y. J. Lin, K. Jim\'enez-Garc\'i, and I. B. Spielman, Spin–orbit-coupled Bose–Einstein condensates, \href{https://www.nature.com/articles/nature09887}{Nature {\bf 471}, 83 (2011)}.

\bibitem{Galitski2013} 
V. Galitski and I. B. Spielman, Spin-orbit coupling in quantum gases, \href{https://www.nature.com/articles/nature11841}{Nature (London) {\bf 494}, 49 (2013)}.

\bibitem{Wu2016}
Z. Wu, L. Zhang, W. Sun, X. Xu, B. Wang, S. Ji, Y Deng, S. Chen, X. Liu, and J. Pan,
Realization of two-dimensional spin-orbit coupling for Bose-Einstein condensates, \href{https://www.science.org/doi/abs/10.1126/science.aaf6689}{Science {\bf 354}, 83 (2016)}.

\bibitem{Sun2018}
W. Sun, B. Wang, X. Xu, C. Yi, L. Zhang, Z. Wu, Y. Deng, X. Liu, S. Chen, and J. Pan
Highly controllable and robust 2D Spin-Orbit Coupling for quantum gases, \href{https://journals.aps.org/prl/abstract/10.1103/PhysRevLett.121.150401}{Phys. Rev. Lett. {\bf 121}, 150401 (2018)}.

\bibitem{Wang2021}
Z. Wang, X. Cheng, B. Wang, J. Zhang, Y. Lu, C. Yi, S. Niu, Y. Deng, X. Liu, S. Chen, and J. Pan,
Realization of an ideal Weyl semimetal band in a quantum gas with 3D spin-orbit coupling, \href{https://www.science.org/doi/abs/10.1126/science.abc0105}{Science {\bf 372}, 271 (2021)}.

\bibitem{Hamner2015}
C. Hamner, Y. Zhang, M. A. Khamehchi, M. J. Davis,
and P. Engels, Spin-orbit-coupled Bose-Einstein condensates in a one-dimensional optical lattice, \href{https://link.aps.org/doi/10.1103/PhysRevLett.114.070401}{Phys. Rev. Lett. {\bf~114}, 070401 (2015)}.

\bibitem{Hamner2014}
C. Hamner, C. Qu, Y. Zhang, J. J. Chang, M. Gong, C. Zhang, and P. Engels, Dicke-type phase transition in a spin-orbit-coupled Bose–Einstein condensate, \href{https://www.nature.com/articles/ncomms5023}{Nat. Commun. {\bf~5}, 4023 (2014)}.

\bibitem{Bergmann1998}
K. Bergmann, H. Theuer, and B. W. Shore, Coherent population transfer among quantum states of atoms and molecules, \href{https://journals.aps.org/rmp/abstract/10.1103/RevModPhys.70.1003}{Rev. Mod. Phys. {\bf~70}, 1003 (1998)}.

\bibitem{Vitanov2017}
N. V. Vitanov, A. A. Rangelov, B. W. Shore, and K. Bergmann, Stimulated Raman adiabatic passage in physics, chemistry, and beyond, \href{https://journals.aps.org/rmp/abstract/10.1103/RevModPhys.89.015006}{Rev. Mod. Phys. {\bf~89}, 015006 (2017)}.

\bibitem{Bergmann2019}
K. Bergmann, H. C. N{\"a}gerl, C. Panda, G. Gabrielse, E. Miloglyadov, M. Quack, G. Seyfang, G. Wichmann, S. Ospelkaus, A. Kuhn, S. Longhi, A. Szameit, P. Pirro1, B. Hillebrands, X. Zhu, J. Zhu, M. Drewsen, W. K. Hensinger, S. Weidt, T. Halfmann, H. Wang, G. S. Paraoanu, N. V. Vitanov, J. Mompart, T. Busch, T. J. Barnum, D. D. Grimes, R. W. Field, M. G. Raizen, E. Narevicius, M. Auzinsh, D. Budker, A. P{\'a}lffy, and C. H. Keitel, Roadmap on STIRAP applications, \href{https://iopscience.iop.org/article/10.1088/1361-6455/ab3995/meta}{J. Phys. B: At. Mol. Opt. Phys. {\bf~52}, 202001 (2019)}. 


\bibitem{Eckert2006}
K. Eckert, J. Mompart, R. Corbal{\'a}n, M. Lewenstein, and G. Birkl, Three level atom optics in dipole traps and waveguides, \href{https://www.sciencedirect.com/science/article/pii/S003040180600486X?casa_token=Wqfv5o_ejUcAAAAA:yKBoMLf6XKGcNBvgrWR7A7nzCCFq8ogc9G8GeAmoM--Di1DM0Az85B9R761xKNH_boSGJWU7qMs}{ Opt. Commun. {\bf~264}, 264 (2006)}.

\bibitem{Graefe2006}
E. M. Graefe, H. J. Korsch, and D. Witthaut, Mean-field dynamics of a Bose-Einstein condensate in a time-dependent triple-well trap: Nonlinear eigenstates, Landau-Zener models, and stimulated Raman adiabatic passage, \href{https://journals.aps.org/pra/abstract/10.1103/PhysRevA.73.013617}{Phys. Rev. A {\bf 73}, 013617 (2006)}.

\bibitem{Longhi2019}
S. Longhi, Topological pumping of edge states via adiabatic passage, \href{https://journals.aps.org/prb/abstract/10.1103/PhysRevB.99.155150}{Phys. Rev. B {\bf~99}, 155150 (2019)}.

\bibitem{Lobanov2014}
V. E. Lobanov, Y. V. Kartashov, and V. V. Konotop, Fundamental, Multipole, and semi-vortex Gap Solitons in Spin-Orbit Coupled Bose-Einstein Condensates, \href{https://journals.aps.org/prl/abstract/10.1103/PhysRevLett.112.180403}{Phys. Rev. Lett. {\bf 112}, 180403 (2014)}.

\bibitem{Malomed2019} B. A. Malomed, Vortex solitons: Old results and new perspectives, Physica D {\bf 399}, 108 (2019).

\bibitem{experiment3}
K. Jim\'enez-Garc\'{\i}a,  and L. J. LeBlanc, R. A. Williams, M. C. Beeler, C. Qu and M. Gong, and C. Zhang, and  I. B. Spielman, Tunable spin-orbit coupling via strong driving in ultracold-atom systems \href{https://journals.aps.org/prl/abstract/10.1103/PhysRevLett.114.125301}{Phys. Rev. Lett. {\bf 114}, 125301 (2015)}.

\bibitem{Zhang2013_1}
Y. Zhang, G. Chen, and C. Zhang, Tunable spin–orbit coupling and quantum phase transition in a trapped Bose-Einstein condensate, \href{https://www.nature.com/articles/srep01937}{Sci. Rep. {\bf 3}, 1937 (2013)}.

\bibitem{Chin2010}
C Chin, R Grimm, P Julienne, and E Tiesinga, Feshbach Resonances in Ultracold Gases, \href{https://journals.aps.org/rmp/abstract/10.1103/RevModPhys.82.1225}{Rev. Mod. Phys. {\bf 82}, 1225 (2010)}.

\bibitem{Hui2017}
H. Hui, Y. Zhang, C. Zhang, and V. W. Scarola, Superfluidity in the absence of kinetics in spin-orbit-coupled optical lattices, \href{https://journals.aps.org/pra/abstract/10.1103/PhysRevA.95.033603}{Phys. Rev. A {\bf 95}, 033603 (2017)}. 

\bibitem{Zhang2013}
Y. Zhang and C. Zhang, Bose-Einstein condensates in spin-orbit-coupled optical lattices: flat bands and superfluidity, \href{https://journals.aps.org/pra/abstract/10.1103/PhysRevA.87.023611}{Phys. Rev. A {\bf~87}, 023611 (2013)}. 

\bibitem{Zhang2015}
Y. Zhang, Y. Xu, and T. Busch, Gap solitons in spin-orbit-coupled Bose-Einstein condensates in optical lattices, \href{https://journals.aps.org/pra/abstract/10.1103/PhysRevA.91.043629}{Phys. Rev. A {\bf~91}, 043629 (2015)}.

\bibitem{Kartashov2016a} 
Y. V. Kartashov, V. V. Konotop, D. A. Zezyulin, and L. Torner, Bloch oscillations in optical and Zeeman lattices in the presence of spin-orbit coupling, \href{https://journals.aps.org/prl/abstract/10.1103/PhysRevLett.117.215301}{Phys. Rev. Lett. {\bf 117}, 215301 (2016)}.

\bibitem{Kartashov2016b} 
Y. V. Kartashov, V. V. Konotop, D. A. Zezyulin, and L. Torner, Dynamic localization in optical and Zeeman lattices in the presence of spin-orbit coupling, \href{https://journals.aps.org/pra/abstract/10.1103/PhysRevA.94.063606}{Phys. Rev. A {\bf 94},  063606 (2016)}.

 

\bibitem{Kohn1959} 
W. Kohn, Analytic properties of Bloch waves and Wannier functions, \href{https://journals.aps.org/pr/abstract/10.1103/PhysRev.115.809}{Phys. Rev. {\bf 115}, 809 (1959)}. 

 

\bibitem{McEndoo2010}
S. McEndoo, S. Croke, J. Brophy, and Th. Busch, Phase evolution in spatial dark states, \href{https://journals.aps.org/pra/abstract/10.1103/PhysRevA.81.043640}{Phys. Rev. A {\bf 81}, 043640 (2010)}.

\bibitem{Greentree2004}
A. D. Greentree, J. H. Cole, A. R. Hamilton, and L. C. L. Hollenberg, Coherent electronic transfer in quantum dot systems using adiabatic passage, \href{https://journals.aps.org/prb/abstract/10.1103/PhysRevB.70.235317}{Phys. Rev. B {\bf 70}, 235317 (2004)}.

\bibitem{Merkel2007} 
W. Merkel, H. Mack, M. Freyberger, V. V. Kozlov, W. P. Schleich, and B. W. Shore, Coherent transport of single atoms in optical lattices, \href{https://journals.aps.org/pra/abstract/10.1103/PhysRevA.75.033420}{Phys. Rev. A {\bf 75}, 033420 (2007)}.

\bibitem{Nesterenko2009}
V. O. Nesterenko, A. N. Novikov, F. F. de Souza Cruz, and E. L.  Lapolli, STIRAP transport of Bose-Einstein condensate in triple-well trap, \href{https://link.springer.com/article/10.1134/S1054660X09040148}{Laser physics {\bf 19}, 616 (2009)}.

\bibitem{Marzari2012}
N. Marzari, A. A. Mostofi, J. R. Yates, I. Souza, and D. Vanderbilt, Maximally localized Wannier functions: Theory and applications, \href{https://journals.aps.org/rmp/abstract/10.1103/RevModPhys.84.1419}{Rev. Mod. Phys. {\bf 84}, 1419 (2012)}.

\bibitem{Salgueiro2009}
J. R. Salgueiro, M. Zacar{\'e}s, H. Michinel, and A. Ferrando, Vortex replication in Bose-Einstein condensates trapped in double-well potentials, \href{https://journals.aps.org/pra/abstract/10.1103/PhysRevA.79.033625}{Phys. Rev. A {\bf 81}, 043640 (2009)}.

\end{thebibliography}
\end{document}